\documentclass[%
 reprint,
 amsmath,amssymb,
 aps,
]{revtex4-1}
\usepackage{graphics}
\usepackage{graphicx}
\usepackage{dcolumn}
\usepackage{bm}

\begin{document}

\preprint{APS/123-QED}

\title{A global estimation of the lower bound of the privacy amplification term for
decoy-state quantum key distribution
 }
\author{Haodong Jiang}
\affiliation{State Key Laboratory of Mathematical Engineering and Advanced Computing, Zhengzhou, Henan, China }
\author{Ming Gao}
\email{gaoming.zhengzhou@gmail.com}
\affiliation{State Key Laboratory of Mathematical Engineering and Advanced Computing, Zhengzhou, Henan, China }
\author{Hong Wang}
\affiliation{State Key Laboratory of Mathematical Engineering and Advanced Computing, Zhengzhou, Henan, China }
\author{Hongxin Li}
\affiliation{State Key Laboratory of Mathematical Engineering and Advanced Computing, Zhengzhou, Henan, China }
\author{Zhi Ma}
\email{ma\_zhi@163.com}
\affiliation{State Key Laboratory of Mathematical Engineering and Advanced Computing, Zhengzhou, Henan, China }

\date{\today}

\begin{abstract}
The privacy amplification term, of which the lower bound needs to be estimated with the decoy-state method, plays a positive role in the secure key rate formula for decoy-state quantum key distribution.
In previous work, the yield and the bit error rate of single-photon state are estimated separately to gain this lower bound.
In this work, we for the first time take the privacy amplification term as a whole to consider this lower bound. The mathematical description for the correlation between the yield and the bit error rate of single-photon state is given with just two unknown variables.
Based on this, we obtain the global estimation of this lower bound for both BB84 protocol and measurement-device-independent protocol.
The results of numerical simulation show that the global estimation can significantly improve the performance of quantum key distribution.
\begin{description}
\item[PACS numbers]
03.67.Dd, 42.81.Gs, 03.67.Hk
\end{description}
\end{abstract}

\pacs{Valid PACS appear here}
\maketitle


\section{\label{sec:level1}Introduction}
Quantum key distribution (QKD) based on the laws of quantum physics can theoretically present an unconditionally secure communication \cite{bennett1984quantum,mayers2001unconditional,ekert1991quantum}.
However, there is a gap between its theory and practice due to the imperfection in real-life implementation. Particularly, the eavesdropper (Eve) can launch attacks aiming at the imperfect single-photon source and the limited detector efficiency in practical QKD system \cite{brassard2000limitations,pns2002quantum,zhao2008quantum,xu2010experimental,weier2011quantum,jain2011device}.
By utilizing the decoy-state method \cite{hwang2003quantum,lo2005decoy,wang2005beating}, the practical QKD setups with an imperfect single-photon source can be still secure.

To deal with the threat coming from the detectors \cite{lydersen2010hacking}, several approaches have been proposed. One is device-independent QKD (DI-QKD) \cite{acin2007device} of which the security is based on the violation of a Bell inequality. However, DI-QKD con not apply to existing practical system because a loophole-free Bell test at the moment is still unavailable. Another one is measurement-device-independent quantum key distribution (MDI-QKD) \cite{braunstein2012side,lo2012measurement} based on the idea of entanglement swapping which can remove all detector side channel attacks.

The security of BB84 protocol with imperfect devices is analyzed in \cite{GLLP2004security,inamori2007unconditional,scarani2008quantum,cai2009finite,lim2014concise}. The security of MDI-QKD protocol is researched in \cite{lo2012measurement,tomamichel2012tight,curty2014finite}. Some useful formulas are given to calculate the secure key rate for practical BB84 protocol and MDI-QKD protocol. The privacy amplification term makes a positive contribution in these secure key rate formulas and it can not be measured in the experiment.
In asymptotic case, the yield of single-photon state is basis independent \cite{wei2013decoy,wang2013three,yu2013decoy}. Then the privacy amplification term can be calculated in just one basis.

In previous work \footnote{For simplicity, the analysis in Sec. \ref{sec:level1} is for BB84 protocol. The same analysis for MDI-QKD protocol is presented in Sec. \ref{sec:level3}.}, the lower bound of this term is obtained by estimating the lower bound of the yield \(Y_1\) of single-photon state and the upper bound of the bit error rate \(e_1\) of single-photon state separately.
The lower bound of the yield \(Y_1\) is estimated from the gain equations while the upper bound of the bit error rate \(e_1\) is estimated from the quantum bit error rate (QBER) equations.
The yield \(Y_i\) of \(i\)-photon state existing in both the gain equations and the QBER equations is the link between the estimation of lower bound of \(Y_1\) and that of upper bound of \(e_1\).
When \(Y_i\) is one certain value, the minimum of \(Y_1\) is reached. But the maximum of \(e_1\) may be reached as \(Y_i\) is another certain value.
That is to say, the lower bound of \(Y_1\) and the upper bound of \(e_{1}\) may not be simultaneously reached.
Thus, the separate estimation can just bring a lower bound of the privacy amplification term instead of the minimum.

Inspired by Wang's method \cite{wang2005beating,wang2013three,yu2013three,zhou2014tightened}, we give a mathematical description of the correlation between \(Y_1\) and \(e_{1}\) with just two unknown variables.
In particular, we will show that globally estimating the lower bound of the privacy amplification term  is equal to finding the minimum of a bivariate continuous function in a closed area. Thus the minimum of the privacy amplification term can be attained with the global estimation and higher secure key rate can be achieved.

The article is organized as follows. Section \ref{sec:level2} introduces the global estimation of the lower bound of the privacy amplification term for BB84 protocol. The global estimation for MDI-QKD protocol will be discussed in section \ref{sec:level3}. We conclude our work in section \ref{sec:level4}.

\section{\label{sec:level2} The global estimation of the lower bound of the privacy amplification term for BB84 protocol}
The privacy amplification term for BB84 protocol is given by \({Y_1}[1 - H({e_1})]\), where $Y_1$ and $e_1$ are, respectively, the yield and the bit error rate of single-photon state. Here in this section, firstly we mathematically characterize the correlation between $Y_1$ and $e_1$. Then the minimum of \({Y_1}[1 - H({e_1})]\) is given with the method of global estimation. Lastly, the numerical simulation is performed to make a comparison in performance of QKD protocol between the global estimation and the separated estimation.
\subsection{\label{sec:level2A} The correlation between $Y_1$ and $e_1$}
Given a weak coherent state source which sends three different kinds of optical pulses with intensities \(\omega\), \(\upsilon\) and \(\mu\) \((0 = \omega  < \upsilon  < \mu )\), the overall gains which mean the probability for Bob to obtain a detection event in one pulse are given by following three equations,
\begin{eqnarray}\label{eq:gains}
{Q_\mu } &&= \sum\limits_{i = 0}^\infty  {{e^{-\mu} }\frac{{{\mu ^i}}}{{i!}}{Y_i}}, \\
{Q_\upsilon } &&= \sum\limits_{i = 0}^\infty  {{e^{-\upsilon} }\frac{{{\upsilon ^i}}}{{i!}}{Y_i}}, \\
&&{Q_\omega } = {Y_0},
\end{eqnarray}
where \(Q_\nu\) and \(Y_i\) are, respectively, the overall gain with intensity \(\nu\) \((\nu \in \{ \omega ,\upsilon ,\mu \} )\) and the yield of \(i\)-photon state.

 We denote \(E_\nu\) to be the overall QBER with intensity \(\nu\), \(e_i\) to be the bit error rate of \(i\)-photon state. The overall QBER equations can be given by
\begin{eqnarray}\label{eq:QBERs}
{E_\mu}{Q_\mu } &&= \sum\limits_{i = 0}^\infty  {{e^{-\mu} }\frac{{{\mu ^i}}}{{i!}}{e_i}{Y_i}}, \\
{E_\upsilon}{Q_\upsilon } &&= \sum\limits_{i = 0}^\infty  {{e^{-\upsilon} }\frac{{{\upsilon ^i}}}{{i!}}{e_i}{Y_i}}, \\
&&{E_\omega }{Q_\omega } ={e_0} {Y_0}.
\end{eqnarray}
It is important to note that \(Y_0\) is equal to the gain \(Q_\omega\) when Alice does not send any optical pulse, which includes the detector dark count and other background  contributions. As the background is random, we assume that \({E_\omega }={e_0}=0.5\).

As three equations can only fix three variables, we temporarily take \(Y_i\) (\(i\ge 3\)) as known variables. Then three gain equations can be solved according to Cramer's rule. \(Y_1\) is given by
\begin{eqnarray}\label{eq:Y1old}
{Y_1} = &&\frac{\mu }{{\upsilon (\mu  - \upsilon )}}({e^\upsilon }{Q_\upsilon } - {Y_0}) - \frac{\upsilon }{{\mu (\mu  - \upsilon )}}({e^\mu }{Q_\mu } - {Y_0}) +\nonumber \\
&&\sum\limits_{i = 3}^\infty  {\frac{{({\mu ^{i - 1}}\upsilon  - {\upsilon ^{i - 1}}\mu )}}{{i!(\mu  - \upsilon )}}} {Y_i}.
\end{eqnarray}

Similarly, \({e_1}{Y_1}\) can be gained by
\begin{eqnarray}\label{eq:e1Y1old}
{e_1}{Y_1}&& = \frac{\mu }{{\upsilon (\mu  - \upsilon )}}({e^\upsilon }{E_\upsilon }{Q_\upsilon } - {e_0}{Y_0}) - \frac{\upsilon }{{\mu (\mu  - \upsilon )}}\nonumber \\
&&({e^\mu }{E_\mu }{Q_\mu } - {e_0}{Y_0}) +
\sum\limits_{i = 3}^\infty  {\frac{{({\mu ^{i - 1}}\upsilon  - {\upsilon ^{i - 1}}\mu )}}{{i!(\mu  - \upsilon )}}{e_i}} {Y_i}.
\end{eqnarray}

From equation (\ref{eq:Y1old}) and equation(\ref{eq:e1Y1old}), we can get that there are infinite variables \(Y_i\) (\(i\ge 3\)) simultaneously influencing the values of \(Y_1\) and \({e_1}{Y_1}\). Then the privacy amplification term is influenced by infinite variables.
It is computationally infeasible to find the minimum of a function with infinite variables.
Fortunately, we find a way to reduce the number of unknown variables to two inspired by Wang's method \cite{wang2005beating,wang2013three}.
We define a state of which the density operator is \(\rho  = \sum\limits_{i = 3}^\infty  {\frac{{({\mu ^{i - 1}}\upsilon  - {\upsilon ^{i - 1}}\mu )}}{{\Omega i!(\mu  - \upsilon )}}} \left| i \right\rangle \left\langle i \right|\) \((\Omega  = \sum\limits_{i = 3}^\infty  {\frac{{({\mu ^{i - 1}}\upsilon  - {\upsilon ^{i - 1}}\mu )}}{{i!(\mu  - \upsilon )}}} >0)\). The yield and the bit error rate of this state can be given by
\begin{eqnarray}
{Y_\rho } &&= \sum\limits_{i = 3}^\infty  {\frac{{({\mu ^{i - 1}}\upsilon  - {\upsilon ^{i - 1}}\mu )}}{{i!(\mu  - \upsilon )\Omega }}} {Y_i},\label{eq:YP}\\
{e_\rho }{Y_\rho } &&= \sum\limits_{i = 3}^\infty  {\frac{{({\mu ^{i - 1}}\upsilon  - {\upsilon ^{i - 1}}\mu )}}{{i!(\mu  - \upsilon )\Omega }}} {e_i}{Y_i}.\label{eq:epYP}
\end{eqnarray}
Then equation (\ref{eq:Y1old}) and equation (\ref{eq:e1Y1old}) can be rewritten as
\begin{eqnarray}
{Y_1} = &&\frac{\mu }{{\upsilon (\mu  - \upsilon )}}({e^\upsilon }{Q_\upsilon } - {Y_0}) - \frac{\upsilon }{{\mu (\mu  - \upsilon )}}({e^\mu }{Q_\mu } - {Y_0}) +\nonumber \\
+{\Omega}{Y_\rho },\label{eq:Y1}\\
{e_1}{Y_1}&& = \frac{\mu }{{\upsilon (\mu  - \upsilon )}}({e^\upsilon }{E_\upsilon }{Q_\upsilon } - {e_0}{Y_0}) - \frac{\upsilon }{{\mu (\mu  - \upsilon )}}\nonumber \\
&&({e^\mu }{E_\mu }{Q_\mu } - {e_0}{Y_0}) +{\Omega}{e_\rho }{Y_\rho }.\label{eq:e1Y1}
\end{eqnarray}

Thus ${Y_1}$ and ${e_1}{Y_1}$ is determined by the gains and the QBERs which can be measured in the experiment except the yield and the bit error rate of state \(\rho\).
State \(\rho\) is the link between the calculations of ${Y_1}$ and ${e_1}{Y_1}$. The yield ${Y_\rho }$ of state \(\rho\) as a unknown variable simultaneously influences the estimations of both ${Y_1}$ and ${e_1}$. In \cite{hayashi2007general}, ${Y_\rho }$ is set to 0 to get the lower bound of ${Y_1}$ while ${e_\rho }$ and ${Y_\rho }$ are both set to 1 to get the upper bound of ${e_1}$. Thus the contradiction that ${Y_\rho }$ cannot be simultaneously 0 and 1 emerges.

The quantity of the privacy amplification term is \({Y_1}[1-H(e_1)]\), which is a bivariate continuous function of ${Y_\rho }$ and ${e_\rho }$. The minimum of the continuous function on the closed area can be attained. This is one reason why we should consider the global lower bound of \({Y_1}[1-H(e_1)]\) instead of calculating the lower bound of ${Y_1}$ and the upper bound of $e_1$ separately. In previous work \cite{lo2005decoy,wang2005beating,ma2005practical,hayashi2007general}, the lower bound of ${Y_1}$ is gained by utilizing the gain equations. In fact, ${Y_1}$ also exists in QBER equations where the information of ${Y_1}$ is not extracted. This is another motivation that the global lower bound of \({Y_1}[1-H(e_1)]\) should be considered.
\subsection{\label{sec:level2B} The global lower bound of \({Y_1}[1-H(e_1)]\)}

According to previous work \cite{lo2005decoy,wang2005beating,ma2005practical,hayashi2007general}, the most accurate estimations of \(Y_1\) and \(e_1\) are given by
\begin{eqnarray}
{Y_1}\ge{Y_1^L} = &&\frac{\mu }{{\upsilon (\mu  - \upsilon )}}({e^\upsilon }{Q_\upsilon } -{Y_0})\nonumber\\
&& - \frac{\upsilon }{{\mu (\mu  - \upsilon )}}({e^\mu }{Q_\mu } - {Y_0}) ,\label{eq:Y1L}\\
{e_1}\le{e_1^U} &&= \frac{({e^\upsilon }{E_\upsilon }{Q_\upsilon }- {e_0}{Y_0})}{{\upsilon}{Y_1^L}}.\label{eq:e1Y1U}
\end{eqnarray}
According to the corollary in appendix, the global lower bound of \({Y_1}[1-H(e_1)]\) can be gained by

\begin{eqnarray}\label{eq:globalLBB84}
{Y_1}(1-H&&(e_1))\ge{({Y_1^L}+\theta})[1-H(\frac{{e_1^U}{Y_1^L}}{Y_1^L+\theta})],\nonumber\\
\theta=&&\frac{1}{{\mu (\mu  - \upsilon )}}[\upsilon ({e^\mu }{E_\mu }{Q_\mu } - {e_0}{Y_0})\nonumber\\
&&-\mu ({e^\upsilon }{E_\upsilon }{Q_\upsilon } - {e_0}{Y_0})]>0.
\end{eqnarray}

To make a clear comparison, we denote \((Y_1^G,e_1^G)\) as the point where the minimum is achieved. Corresponding to equation (\ref{eq:Y1L}) and equation (\ref{eq:e1Y1U}), \(Y_1^G\) and \(e_1^G\) are given by

\begin{eqnarray}
{Y_1^G} = {Y_1^L}+\theta,\label{eq:Y1G}\\
{e_1^G} = \frac{{e_1^U}{Y_1^L}}{{Y_1^L}+\theta}.\label{eq:e1G}
\end{eqnarray}

Here \(\theta\) can be considered the information of \(Y_1\) coming from the QBER equations, which is abandoned for the separate estimation. By globally considering the lower bound of the privacy amplification term, we successfully extract it.

\subsection{\label{sec:level2C} Numerical simulation for BB84 protocol}
With the observed gains and error rates, the final secure key rate can be calculated \cite{GLLP2004security} by
\begin{eqnarray}
{R}\ge{p_1^{\mu}}Y_1[1-H(e_1)]-{Q_\mu}fH(E_\mu)\label{eq:key1},
\end{eqnarray}
where \(p_1^{\mu}\) is the probability that Alice sends a single-photon state pulse corresponding to signal state \(\mu\); \(f\) is the error correction inefficiency; \(H(x)=- x{\log _2}(x) - (1 - x){\log _2}(1 - x)\) is the binary Shannon entropy function.
For a fair comparison, we use the same parameters in \cite{yu2013three,zhou2014tightened} summarized in table \ref{tab:table1}. For simplicity, the detection efficiency is put to the overall channel transmission, hence we only need to assume the 100\% detection efficiency at Bob's side.

 \begin{table}[htbp!]
\caption{\label{tab:table1}%
List of parameters for numerical simulation}
\begin{ruledtabular}
\begin{tabular}{cccc}
\textrm{\(e_0\)}&
\textrm{\(f\)}&
\textrm{\({p_d}\)}&
\textrm{\(e_d\)}\\
\colrule
0.5 & 1.16 & \text{$3\times{10^{-6}}$} & \text{$1.5\%$}\\
\end{tabular}
\end{ruledtabular}
\end{table}

\begin{figure}[h]
\centering
\includegraphics[height=0.32\textwidth,width=0.5\textwidth]{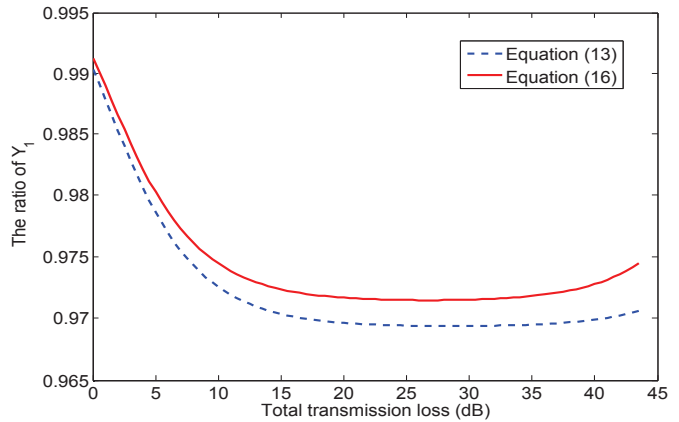}
\caption{\label{fig:Y1}
 (Color online) The ratio of the estimation of \(Y_1\) to the asymptotic limit calculated with the infinite-intensity decoy-state method vs the total channel transmission loss for three-intensity
decoy-state BB84 protocol.  We set \(\upsilon=0.1\), \(\mu=0.5\) for decoy state and signal state, respectively.
}
\end{figure}

\begin{figure}[h]
\centering
\includegraphics[height=0.32\textwidth,width=0.5\textwidth]{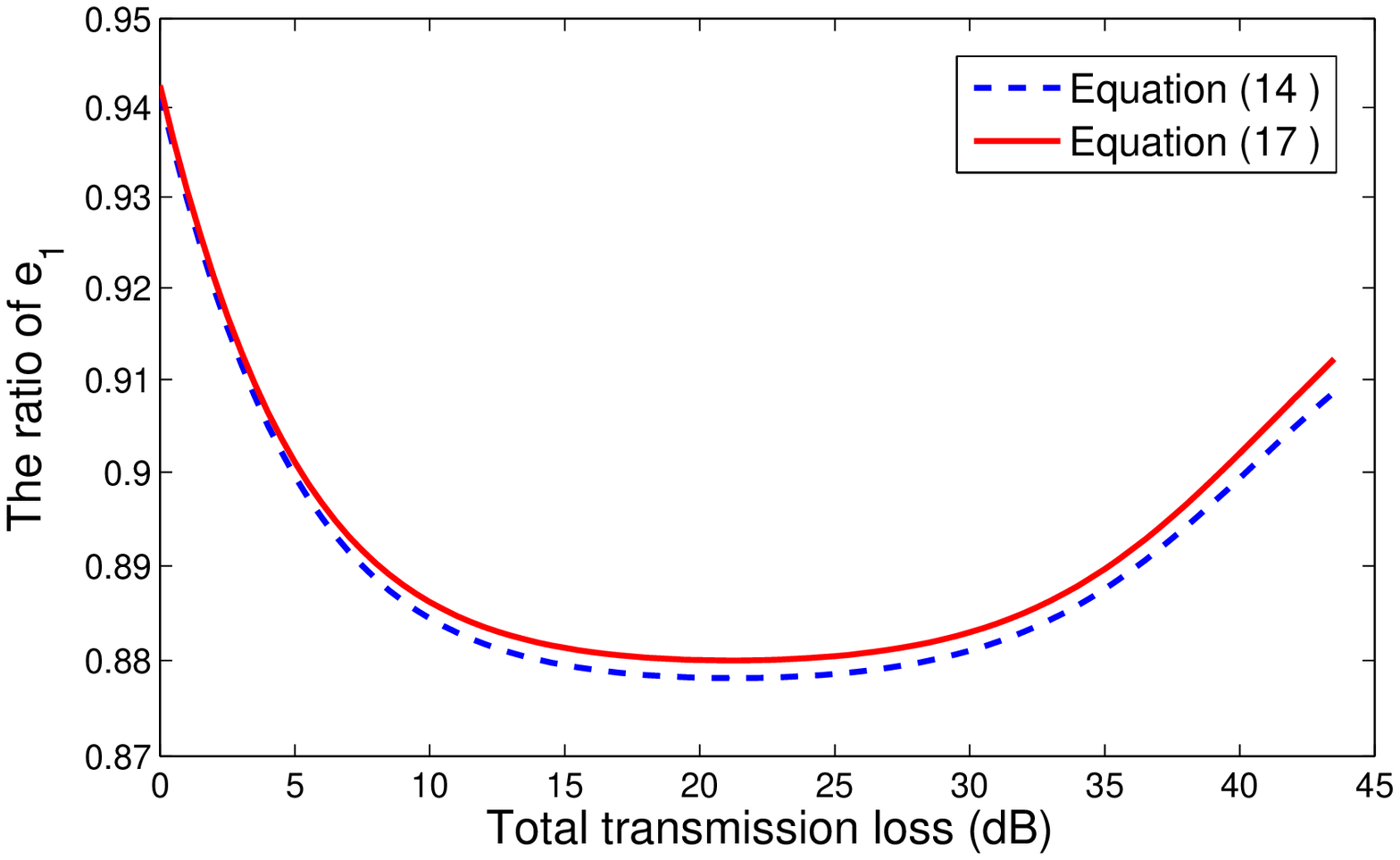}
\caption{\label{fig:e1}
 (Color online) The ratio of the asymptotic limit calculated with the infinite-intensity decoy-state method to the estimation of \(e_1\) vs the total channel transmission loss for three-intensity
decoy-state BB84 protocol.  We set \(\upsilon=0.1\), \(\mu=0.5\) for decoy state and signal state, respectively.
}
\end{figure}

\begin{figure}[h]
\centering
\includegraphics[height=0.35\textwidth,width=0.5\textwidth]{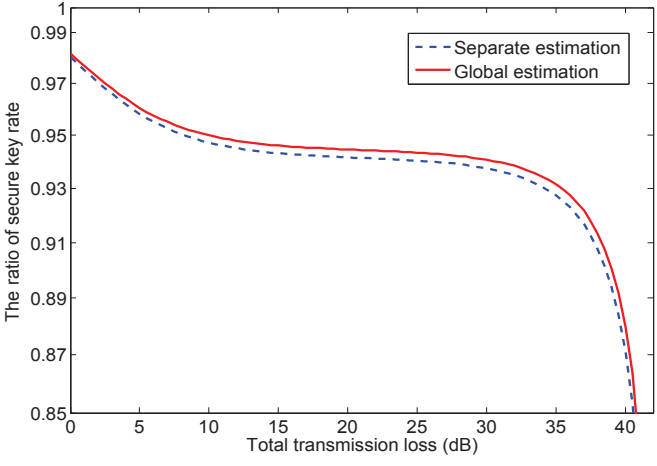}
\caption{\label{fig:key1}
 (Color online) The ratio of the secure key rate calculated with the three-intensity decoy-state method to the asymptotic limit calculated with the infinite-intensity decoy-state method vs the total channel transmission loss for
decoy-state BB84 protocol.  We set \(\upsilon=0.1\), \(\mu=0.5\) for decoy state and signal state, respectively.
}
\end{figure}

The ratios of the estimations of \(Y_1\) with two methods (equation (\ref{eq:Y1L}) and equation (\ref{eq:Y1G})) to the asymptotic limit calculated with the infinite-intensity decoy-state method  are shown in figure \ref{fig:Y1}. The ratios of the asymptotic limit of \(e_1\) to the estimations with two methods (equation (\ref{eq:e1Y1U}) and equation (\ref{eq:e1G})) are shown in figure \ref{fig:e1}. The ratios of the secure key rates computed with two methods (separate estimation and global estimation) to the asymptotic limit are shown in figure \ref{fig:key1}.
From the results, we can see tighter estimations of \(Y_1\)  and \(e_1\) are gained with the method of global estimation. Thus, higher secure key rates are achieved.

\section{\label{sec:level3} The global estimation of the lower bound of the privacy amplification term for MDI-QKD protocol}
For MDI-QKD protocol, the secure key rate is gained \cite{lo2012measurement} by
\begin{eqnarray}\label{eq:MDIsecure key rate}
R \ge p_{11}^zY_{11}^z[1 - H(e_{11}^x)] - Q_{\mu_a \mu_b }^zfH(E_{\mu_a \mu_b }^z),
\end{eqnarray}
where \(p_{11}^z\) is the probability that Alice and Bob simultaneously send single-photon state pulses corresponding to signal state in \(z\) basis; \(Q_{\mu_a \mu_b }^z\) and \(E_{\mu_a \mu_b }^z\) are the gain and QBER when Alice and Bob simultaneously send signal state pulses; \(Y_{11}^z\) and \(e_{11}^x\) are the yield in \(Z\) basis and the bit error rate in \(X\) basis when Alice and Bob simultaneously send single-photon state pulses.

The variable values in (\ref{eq:MDIsecure key rate}) can be measured in the experiment except \(Y_{11}^z\) and \(e_{11}^x\).
So the major task in the calculation of secure key rate is estimating the lower bound of \(Y_{11}^z[1-H(e_{11}^x)]\). In previous work, to get the lower bound of \(Y_{11}^z[1-H(e_{11}^x)]\), the lower bound of \(Y_{11}^z\) and the upper bound of \(e_{11}^x\) are calculated separately.

In fact, \({Y_{11}^z}\) is equal to \(Y_{11}^x\) in asymptotic setting according to \cite{wang2013three}.
As a result, we will not temporarily distinguish the basis of $Y_{11}$ and $e_{11}$.
We will consider the lower bound of \(Y_{11}[1-H(e_{11})]\) as a whole.

 Similarly, in this section we will firstly introduce the mathematical description of the correlation between $Y_{11}$ and $e_{11}$. Then the global lower bound of \({Y_{11}}[1-H(e_{11})]\) is calculated. Lastly, the results of numerical simulation will be given. The following work is on basis of the three-intensity decoy-state MDI-QKD protocol \cite{yu2013three}.

\subsection{\label{sec:level3A} The correlation between $Y_{11}$ and $e_{11}$}
For MDI-QKD protocol, the gain and QBER when Alice (Bob) sends a certain pulse with intensity \(q_a\) (\(q_b\)) can be given by
\begin{eqnarray}
{Q_{{q_a} {q_b} }} &&= \sum\limits_{i,j = 0}^\infty  {{e^{ - ({q_a}  + {q_b} )}}\frac{{{{q_a} ^i}{{q_b} ^j}}}{{i!j!}}} {Y_{ij}},\label{eq:Quv}\\
{E_{{q_a} {q_b} }}{Q_{{q_a} {q_b} }} &&= \sum\limits_{i,j = 0}^\infty  {{e^{ - ({q_a}  + {q_b})}}\frac{{{{q_a} ^i}{{q_b} ^j}}}{{i!j!}}} {e_{ij}}{Y_{ij}},\label{eq:EuvQuv}
\end{eqnarray}
where \(Y_{ij}\) and \(e_{ij}\) is the yield and the bit error rate when Alice (Bob) sends an \(i\)-photon (\(j\)-photon) state pulse.

Given two weak coherent state sources which send three different kinds of optical pulses with intensities \((0={\omega_a}<{\upsilon_a}<{\mu_a})\) and \((0={\omega_b}<{\upsilon_b}<{\mu_b})\), we eliminate the unknown variables \(Y_{0i}\) and \(Y_{j0}\), then get
\begin{eqnarray}\label{eq:Qu1v1}
{e^{(\mu_a  + \mu_b )}}{{\tilde Q}_{\mu_a \mu_b }} = \sum\limits_{i,j = 1}^\infty  {\frac{{{\mu_a ^i}{\mu_b ^j}}}{{i!j!}}} {Y_{ij}},\\
e^{(\mu_a  + \upsilon_b )}}{{\tilde Q}_{\mu_a \upsilon_b }} = \sum\limits_{i,j = 1}^\infty  {\frac{{{\mu_a ^i}{\upsilon_b ^j}}}{{i!j!}}} {Y_{ij},\label{eq:Quv1:2}\\
e^{(\upsilon_a  + \mu_b )}}{{\tilde Q}_{\upsilon_a  \mu_b }} = \sum\limits_{i,j = 1}^\infty  {\frac{{{\upsilon_a  ^i}{\mu_b ^j}}}{{i!j!}}} {Y_{ij},\label{eq:Quv1:3}\\
e^{(\upsilon_a  + \upsilon_b )}}{{\tilde Q}_{\upsilon_a  \upsilon_b }} = \sum\limits_{i,j = 1}^\infty  {\frac{{{\upsilon_a  ^i}{\upsilon_b ^j}}}{{i!j!}}} {Y_{ij},\label{eq:Quv1:4}
\end{eqnarray}
where \({{\tilde Q}_{{\mu _1},{\mu _2}}}({\mu _1} \in \{ \mu_a ,\upsilon_a \},{\mu _2}\in\{\mu_b ,\upsilon_b \} )\) is achieved by
\begin{eqnarray}\label{eq:Qu1v12}
{{\tilde Q}_{{\mu _1}{\mu _2}}} =&& {Q_{{\mu _1}{\mu _2}}} + {e^{ - ({\mu _1} + {\mu _2})}}{Q_{\omega_a \omega_b }} \nonumber\\
 &&-{e^{ - {\mu _1}}}{Q_{\omega_a {\mu _2}}} - {e^{ - {\mu _2}}}{Q_{{\mu _1}\omega_b }}.
\end{eqnarray}

According to \cite{yu2013three}, \(Y_{11}\) can be solved from equations
(\ref{eq:Quv1:2}, \ref{eq:Quv1:3} and \ref{eq:Quv1:4}),
\begin{eqnarray}\label{eq:Y11}
&&{Y_{1,1}} ={Y_{11}^L}+ \sum\limits_{(i + j) \ge 4} {\frac{{{\Upsilon _{i,j}}{Y_{i,j}}}}{{i!j!({\mu _a} - {\upsilon _a})({\mu _b} - {\upsilon _b})}}},\\
&&{\Upsilon _{i,j}} = \upsilon _a^{i - 1}\mu _b^{j - 1}{\upsilon _b}({\mu _a} - {\upsilon _a}) + \mu _a^{i - 1}\upsilon _b^{j - 1}{\upsilon _a}({\mu _b} - {\upsilon _b}) \label{Y11}\nonumber\\
&&- \upsilon _a^{i - 1}\upsilon _b^{j - 1}({\mu _a}{\mu _b} - {\upsilon _a}{\upsilon _b})>0,\nonumber\\
&&{Y_{11}^L}=\frac{1}{{({\mu _a} - {\upsilon _a})({\mu _b} - {\upsilon _b})}}(\frac{{{e^{({\upsilon _a} + {\upsilon _b})}}({\mu _a}{\mu _b} - {\upsilon _a}{\upsilon _b})}}{{{\upsilon _a}{\upsilon _b}}}{\tilde Q_{{\upsilon _a}{\upsilon_b}}}- \nonumber\\
&& \frac{{{e^{({\mu _a} + {\upsilon _b})}}{\upsilon _a}({\mu _b} - {\upsilon _b})}}{{{\mu _a}{\upsilon _b}}}{\tilde Q_{{\mu _a}{\upsilon _b}}}
 - \frac{{{e^{({\upsilon _a} + {\mu _b})}}{\upsilon _b}({\mu _a} - {\upsilon _a})}}{{{\upsilon _a}{\mu _b}}} {\tilde Q_{{\upsilon _a}{\mu _b}}}).\nonumber\\\label{Y11L}
\end{eqnarray}

Similarly, \(e_{11}\) can be solved from the corresponding QBER equations,
\begin{eqnarray}\label{eq:e11}
&&{e_{11}}{Y_{11}}  =  ({e_{11}}{Y_{11}})^L+ \sum\limits_{(i + j) \ge 4} {\frac{{{e_{i,j}}{\Upsilon _{i,j}}{Y_{i,j}}}}{{i!j!({\mu _a} - {\upsilon _a})({\mu _b} - {\upsilon _b})}}},\\
&&({e_{11}}{Y_{11}})^L=\frac{1}{{({\mu _a} - {\upsilon _a})({\mu _b} - {\upsilon _b})}}(\frac{{{e^{({\upsilon _a} + {\upsilon _b})}}({\mu _a}{\mu _b} - {\upsilon _a}{\upsilon _b})}}{{{\upsilon _a}{\upsilon _b}}}\nonumber\\
&&{\tilde Q_{{\upsilon _a}{\upsilon_b}}}{\tilde E_{{\upsilon _a}{\upsilon_b}}} - \frac{{{e^{({\mu _a} + {\upsilon _b})}}{\upsilon _a}({\mu _b} - {\upsilon _b})}}{{{\mu _a}{\upsilon _b}}}{\tilde E_{{\mu _a}{\upsilon _b}}}{\tilde Q_{{\mu _a}{\upsilon _b}}}
 -  \nonumber\\
 &&\frac{{{e^{({\upsilon _a} + {\mu _b})}}{\upsilon _b}({\mu _a} - {\upsilon _a})}}{{{\upsilon _a}{\mu _b}}}{\tilde E_{{\upsilon _a}{\mu _b}}}{\tilde Q_{{\upsilon _a}{\mu _b}}}).
\end{eqnarray}
\({{\tilde E}_{{\mu _1},{\mu _2}}}{{\tilde Q}_{{\mu _1},{\mu _2}}}({\mu _1} \in \{ \mu_a ,\upsilon_a \},{\mu _2}\in\{\mu_b ,\upsilon_b \} )\) is achieved by
\begin{eqnarray}\label{eq:EQu1v12}
{{\tilde E}_{{\mu _1}{\mu _2}}}{\tilde Q}_{{\mu _1}{\mu _2}}=&& {E_{{\mu _1}{\mu _2}}}{Q_{{\mu _1}{\mu _2}}} + {e^{ - ({\mu _1} + {\mu _2})}}{E_{\omega_a \omega_b }}{Q_{\omega_a \omega_b }}- \nonumber\\
 {e^{ - {\mu _1}}}&&{E_{\omega_a {\mu _2}}}{Q_{\omega_a {\mu _2}}} - {e^{ - {\mu _2}}}{E_{{\mu _1}\omega_b }}{Q_{{\mu _1}\omega_b }}.
\end{eqnarray}

It is easy to verify that \({\Upsilon _{i,j}}\) is positive when \((i + j) \ge 4\). So we can define a state of which the density operator is \(\psi= \sum\limits_{\scriptstyle(i+j)\ge4\hfill} {\frac{{{\Upsilon _{i,j}}}}{{i!j!{{(\mu_a - \upsilon_a)(\mu_b - \upsilon_b)}}\Pi}}}{(\left| i \right\rangle \left\langle i \right| \otimes \left| j \right\rangle \left\langle j \right|)}\),
where \(\Pi\) is equal to \(\sum\limits_{\scriptstyle(i+j)\ge4\hfill} {\frac{{{\Upsilon _{i,j}}}}{{i!j!{{(\mu_a  - \upsilon_a )(\mu_b - \upsilon_b)}}}}}\).

Then equation (\ref{eq:Y11}) and equation (\ref{eq:e11}) can be rewritten
\begin{eqnarray}
{Y_{11}} = &&{Y_{11}^L}+\Pi{Y_\psi},\label{eq:Y11our}\\
{e_{11}}{Y_{11}} = &&({e_{11}{Y_{11}^L}})^L+ \Pi{e_\psi}{Y_\psi},\label{eq:e11our}
\end{eqnarray}
where \({Y_\psi}\) and \({e_\psi}\) is the yield and the bit error rate of state \(\psi\).

Thus \(Y_{11}\) and \(e_{11}\) is linked by the state \(\psi\).  \({Y_{11}}(1-H(e_{11}))\) is a bivariate continuous function with two parameter variables
\({Y_\psi}\) and \({e_\psi}\). The lower bound of \(Y_{11}\) can be gained by setting \({Y_\psi}\) to 0 while the upper bound of \(e_{11}\) can be gained by setting \({Y_\psi}\) and \({e_\psi}\) to 1. Thus the lower bound of \({Y_{11}}(1-H(e_{11}))\) can not be reached with the separate estimation. The minimum of \({Y_{11}}(1-H(e_{11}))\) can be attained with the global estimation.

\subsection{\label{sec:level3B} The global lower bound of \({Y_{11}}(1-H(e_{11}))\)}

In \cite{yu2013three}, the lower bound of \({Y_{11}}\) is given in equation (\ref{Y11L}) by setting the last term in equation (\ref{eq:Y11}) to 0.
The upper bound of \(e_{11}\) is given by setting the term \({e_{ij}}{Y_{ij}}\) \((i+j)\ge2\) of \({{\tilde E}_{{\upsilon_a}{\upsilon_b}}}{\tilde Q}_{{\upsilon_a}{\upsilon_b}}\) to 0,
\begin{eqnarray}\label{e11U}
{e_{11}}\le{e_{11}^U} = \frac{{{e^{\upsilon_a+\upsilon_b }}{{\tilde E}_{\upsilon_a \upsilon_b }}{{\tilde Q}_{\upsilon_a \upsilon_b }}}}{{{\upsilon_a}{\upsilon_b}{Y_{11}^{L}}}}.
\end{eqnarray}

According equations (\ref{eq:Y11our}, \ref{eq:e11our} and \ref{e11U}) and corollary in appendix, the global lower bound of \({Y_{11}}(1-H[e_{11}])\) is given by
\begin{eqnarray}\label{eq:Y11e11globallowerbound}
{Y_{11}}[1-H(e_{11})]&&\ge{({Y_{11}^L}+\delta)[1-H(\frac{{e_{11}^U}{Y_{11}^L}}{{Y_{11}^L+\delta}})]},\\
\delta&&={e_{11}^U}{Y_{11}^L}-({e_{11}{Y_{11}}})^L>0\nonumber.
\end{eqnarray}

To make a clear comparison, we denote \((Y_{11}^G,e_{11}^G)\) as the point where the minimum is attained. Corresponding to equation (\ref{Y11L}) and equation (\ref{e11U}), \(Y_{11}^G\) and \(e_{11}^G\) is given by
\begin{eqnarray}
{Y_{11}^G} = {Y_{11}^L}+\delta,\label{eq:Y11G}\\
{e_{11}^G} = \frac{{e_{11}^U}{Y_{11}^L}}{{Y_{11}^L}+\delta}.\label{eq:e11G}
\end{eqnarray}

\subsection{\label{sec:level3C} Numerical simulation for MDI-QKD protocol}
\begin{figure}[h]
\centering
\includegraphics[height=0.32\textwidth,width=0.5\textwidth]{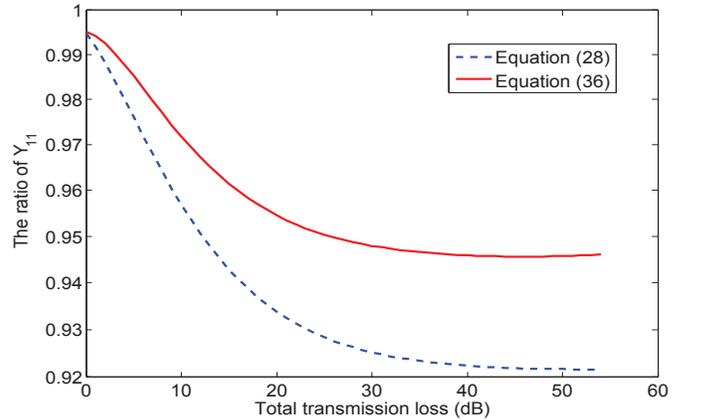}
\caption{\label{fig:Y11}
 (Color online) The ratio of the estimation of \(Y_{11}\) to the asymptotic limit calculated with the infinite-intensity decoy-state method vs the total channel transmission loss for three-intensity
decoy-state MDI-QKD protocol.  We set \({\upsilon}_a={\upsilon}_b=0.1\), \({\mu}_a={\mu}_b=0.5\) for decoy states and signal states, respectively.
}
\end{figure}

\begin{figure}[h]
\centering
\includegraphics[height=0.32\textwidth,width=0.5\textwidth]{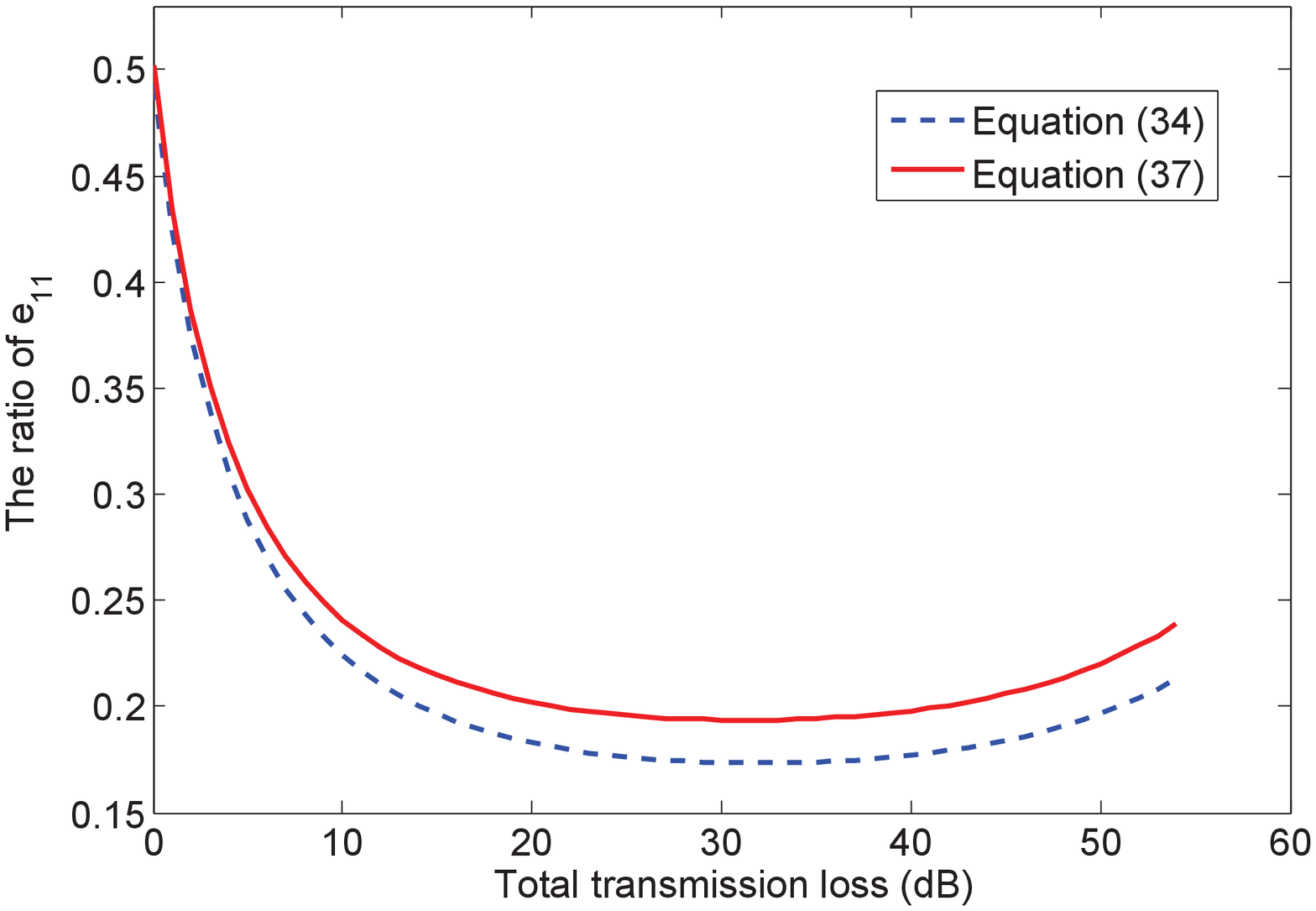}
\caption{\label{fig:e11}
 (Color online) The ratio of the asymptotic limit of \(e_{11}\) calculated with the infinite-intensity decoy-state method to the estimation vs the total channel transmission loss for three-intensity
decoy-state MDI-QKD protocol.  We set \({\upsilon}_a={\upsilon}_b=0.1\), \({\mu}_a={\mu}_b=0.5\) for decoy states and signal states, respectively.}
\end{figure}

\begin{figure}[h]
\centering
\includegraphics[height=0.34\textwidth,width=0.5\textwidth]{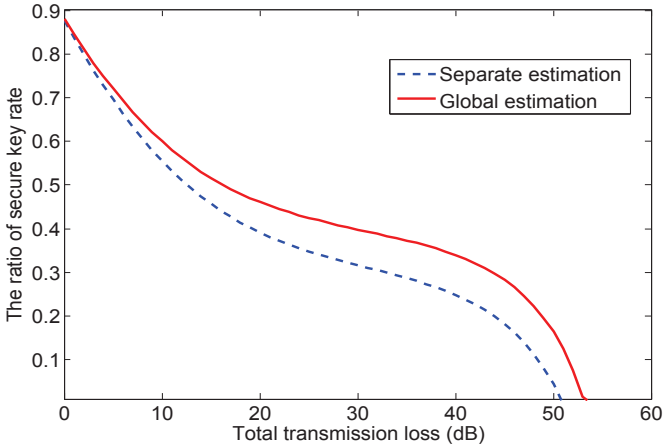}
\caption{\label{fig:key2}
 (Color online) The ratio of secure key rate calculated with the three-intensity decoy-state method to the asymptotic limit calculated with the infinite-intensity decoy-state method vs the total channel transmission loss for decoy-state MDI-QKD protocol.  We set \({\upsilon}_a={\upsilon}_b=0.1\), \({\mu}_a={\mu}_b=0.5\) for decoy states and signal states, respectively.
}
\end{figure}

Numerical simulations are performed with the parameters in table \ref{tab:table1}.
The ratios of the estimations of \(Y_{11}\) with two methods (equation (\ref{Y11L}) and equation (\ref{eq:Y11G})) to the asymptotic limit obtained with the infinite-intensity decoy-state method are shown in figure \ref{fig:Y11}. The ratios of the asymptotic limit of \(e_{11}\) to the estimations with two methods (equation (\ref{e11U}) and equation (\ref{eq:e11G})) are shown in figure \ref{fig:e11}. The ratios of the secure key rates calculated with two methods (separate estimation and global estimation) to the asymptotic limit are shown in figure \ref{fig:key2}.
From the results, we can see tighter estimations of \(Y_{11}\)  and \(e_{11}\) are gained with global estimation. Thus, higher secure key rates are reached.

\section{\label{sec:level4}Conclusion}
The global estimations of the privacy amplification term for both BB84 protocol and MDI-QKD protocol have been researched in this paper. Conventional separate estimation will abandon the information of the yield of single-photon state in QBER equations. With the global estimation of the privacy amplification term, this information has been extracted and the minimum of the privacy amplification term is achieved. Compared with separate consideration, more accurate estimations of the yield and the bit error rate of single-photon state are gained, which thus significantly improve the performance of the quantum key distribution for both BB84 protocol and MDI-QKD protocol.
Additionally, more accurate separate estimation will contribute to more
smaller domain of the bivariate function which thus can further help to obtain a tighter global estimation.
\appendix

\section*{appendix}
\textit{\textbf{Theorem:}} For the bivariate continuous function \(f(x,y) = (A + Cy)[1 - H(\frac{{B + Cxy}}{{A + Cy}})]\) \((A>0,C>0)\) with the definition domain \(\{ (x,y):0 \le x \le 1,0 \le y \le 1,{\frac{{B + Cxy}}{{A + Cy}}<0.5}\} \), the minimum can be attained on the border.

\textit{\textbf{proof:}} Firstly, the partial derivatives of function \(f(x,y)\) are given by
\begin{equation}\label{eq:fx}
{f_x} =  - (A + Cy)H(\frac{{B + Cxy}}{{(A + Cy)}})'\frac{{Cy}}{{A + Cy}},
\end{equation}

\begin{eqnarray}\label{eq:fy}
{f_y} = &&C[1 - H(\frac{{B + Cxy}}{{A + Cy}})] - (A + Cy)H(\frac{{B + Cxy}}{{A + Cy}})'\nonumber\\
&&\frac{{(ACx - BC)}}{{{{(A + Cy)}^2}}}.
\end{eqnarray}
If there is an extreme point \((x_0,y_0)\) \((0<x_0<1,0<y_0<1)\), then \(H(\frac{{B + Cxy}}{{A + Cy}})'\) has to be 0 from the restrict \({f_x} =0\). Combine the restrict \({f_y} =0\), we can get \(C[1 - H(\frac{{B + Cxy}}{{A + Cy}})]=0\). This is in contradiction with our initial assumption.

Function \(f(x,y)\) for a fixed \(y\) is a decreasing function with parameter variable \(x\). So the minimum can be reached where \(x\) is 1. So this problem is converted to searching the minimum of
univariate continuous function \(g(y) = (A + y)[1 - H(\frac{{B + y}}{{A + y}})](0 \le y \le C)\). Calculating the derivative function of \(g(y)\), we can find
\begin{eqnarray}\label{eq:gy}
{g_y}&& = 1 - H(\frac{{B + y}}{{A + y}}) - (A + y)H(\frac{{B + y}}{{A + y}})'\frac{{(A - B)}}{{{{(A + y)}^2}}}\nonumber\\
 &&= 1 + (\frac{{B + y}}{{A + y}})\log (\frac{{B + y}}{{A + y}}) + (\frac{{A - B}}{{A + y}})\log (\frac{{A - B}}{{A + y}})\nonumber\\
 && - (\frac{{A - B}}{{A + y}})\log (\frac{{A - B}}{{B + y}})\nonumber\\
&&= 1 + \log (\frac{{B + y}}{{A + y}}).
\end{eqnarray}

As we assume \(\frac{{B + y}}{{A + y}} < 1/2\), then \({g_y}<0\). That is to say, \({g_y}\) is a decreasing function with parameter variable \(y\).

\textit{\textbf{Corollary:}} For the bivariate continuous function \(f(x,y) = (A + Cy)[1 - H(\frac{{B + Cxy}}{{A + Cy}})]\) \((A>0,C>0)\) with the definition domain \(\{ (x,y):0 \le x \le 1,0 \le y \le 1,{\frac{{B + Cxy}}{{A + Cy}}<0.5}, {(B + Cxy)}<D, {(A+Cy)}>E\}\), the nonzero minimum can be obtained in the following three cases.

case 1: when \((D-B)<C\) and \((D-B)>(E-A)\),
the minimum is \(f(1,\frac{{D - B}}{C})=(A+D-B)[1-H(\frac{D}{A+D-B})]\).

case 2: when \((D-B)<C\) and \((D-B)<(E-A)\),
the minimum is \(f(\frac{D-B}{E-A},\frac{{E- A}}{C})=E[1-H(\frac{D}{E})]\).

case 3: when \((D-B)>=C\),
the minimum is \(f(1,1)=(A + C)[1 - H(\frac{{B + C}}{{A + C}})]\).

\textit{\textbf{proof:}} If we set \({(B + Cxy)}=D\), the function \(f(x,y)\) is converted to an univariate continuous increasing function \((A + Cy)[1 - H(\frac{D}{{A + Cy}})]\). Then it is easy to verify the correctness of corollary combining with the proof of theorem.
\section*{Acknowledgements}
This work is supported by the National High Technology Research and Development Program of China
Grant No.2011AA010803, the National Natural Science Foundation of China Grants No.61472446 and No.U1204602 and the Open
Project Program of the State Key Laboratory of Mathematical Engineering and Advanced Computing Grant No.2013A14.

\bibliography{globalestimation}
\bibliographystyle{unsrt}
\end{document}